# The third- and fourth-order orbital angular momentum multiplexed amplification with ultra-low differential mode gain

TianJin Wen,[1] Shecheng Gao,[2,5] Wei Li,[3] Jiajing Tu,[2,6] Cheng Du,[3] Ji Zhou,[2] Bin Zhang,[1] Weiping Liu,[2] Zhaohui Li,[1,4,7]

1 State Key Laboratory of Optoelectronic Materials and Technologies and School of Electronics and Information Technology, Sun Yat-sen University, Guangzhou,510275, China
2 Department of Electronic Engineering, College of Information Science and Technology, Jinan University, Guangzhou 510632, China
3 Fiberhome Telecommunication Technologies Co. Ltd., Wuhan 430074, China
4 Southern Laboratory of Ocean Science and Engineering (Guangdong, Zhuhai), Zhuhai 519000, China
5 gaosc825@163.com
6 tujiajing@jnu.edu.cn
7 lzhh88@mail.sysu.edu.cn

**In this Letter, a ring-core erbium-doped fiber (RC-EDF), with two-layer erbium-doped structure, supporting up to the fourth-order orbital angular momentum (OAM) mode is designed and fabricated for OAM mode multiplexed amplification. Using the RC-EDF, the third- and fourth-order OAM modes amplification with ultra-low differential mode gain (DMG) is demonstrated by observing both the modal intensity and phase distribution and measuring the modal gain under the fundamental mode core-pumping. The measured average gain of four modes (*l*=+3, -3, +4, -4) multiplexed amplification is higher than 19dB cover the C-band and the DMG is less than 1dB. Additionally, the gain of two conjugate OAM modes are almost the same under different pump power no matter they are amplified simultaneously or separately.**

In order to break through the limit of transmission capacity of the conventional single mode fiber (SMF)[1,2], few-mode/multi-mode fibers based mode division multiplexing (MDM) technology provides a new physical multiplexing dimension to promote the capacity of the optical fiber communication system [3,4]. Orbital angular momentum (OAM) beam with helical wavefront has unlimited topological charge values and inherent orthogonality theoretically, which indicates that OAM modes transmitted in fiber can also be regarded as a kind of candidate solution for MDM [5]. OAM modes in fiber can be expressed as $OAM_{l,m}$, where $|l|$ is the topological charge, and $m$ is the number of concentric rings in the transverse intensity profile of the mode, which can be explained as a superposition of eigenmodes : $OAM^{\pm}_{\pm l,m} = HE^{even}_{l+1,m} \pm iHE^{odd}_{l+1,m}$ and $OAM^{\mp}_{\pm l,m} = EH^{even}_{l-1,m} \pm iEH^{odd}_{l-1,m}$ [6]. Generally, weakly-coupled ring core fiber (RCF) based OAM multiplexing communication system can simplify and alleviate the complexity of the complex multiple-input multiple-output digital signal processing (MIMO-DSP) used in multi-mode fibers [7-9]. To date, there have been many RCFs based OAM multiplexing transmission system using 4×4 MIMO-DSP reported, such as the demonstration with the length of 10km [8], 50km [9] and 100km [10]. In these transmission experiments, the higher-order modes ($|l|≥2$ or 3) are adopted to multiplex into the RCF, due to its low coupling between the adjacent higher-order mode groups and high intra-mode-group modes degeneracy[11]. However, regarding the OAM mode erbium-doped fiber (EDF), the reported works mostly focus on the OAM modes with $|l|≤2$, including the first-order OAM mode amplification based on air-hole EDF [12-13], the second-order OAM mode amplification based on few-mode EDF [14,15]. In order to exploit the technical benefit of a long-haul OAM multiplexing transmission system, a suitable inline OAM mode amplifier for higher-order modes is essential.

As a multi-mode amplifier, the performance of modal gain equalization is critical because the disequilibrium of each mode amplification may induce system outage[16]. Wang *et al.* have demonstrated that two OAM modes are ($|l|$=1,2) simultaneously amplified in the few-mode EDF with high DMG of > 5dB [15]. To overcome this issue, ring-core erbium-doped fiber (RC-EDF) which can form similar distribution of OAM modes is a feasible solution to achieve the highly coincident of the overlap between pump fundamental mode and signal OAM modes for reducing the DMG. As shown in Fig. 1(a), there is a large difference among the pump mode and signal modes in circle core structure. Although it can also achieve gain equalization by designing a complex erbium ions distribution, the ion diffusion effect and other imperfections during the fabrication process make erbium ion distribution different from the original design[20]. Fortunately, as shown in Fig.1 (b), the pump mode shifts to the same distribution as signal OAM modes in the RC-EDF, which can reduce the complexity of both the erbium-doped design and the fabrication. Although the theoretical analysis show that EDF with a ring-core profile can realize low DMG [17], the gain characteristics of RC-EDF at experiment for higher-order ($|l|$>2) OAM modes-multiplexed amplification have not been discussed and reported.

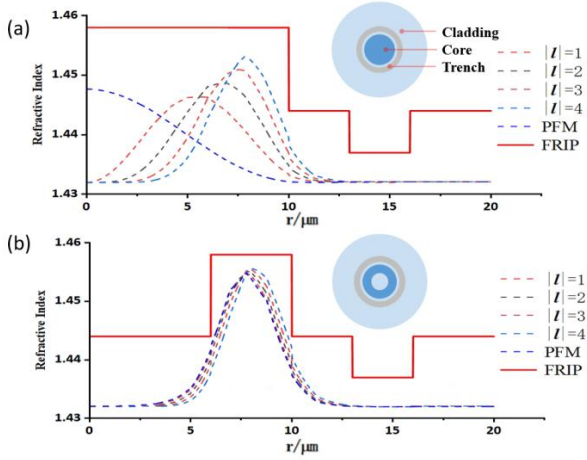

Fig. 1. The FRIP overlaid with the normalized PFM and signal intensity profile of the OAM states $|l|$ = 1, 2, 3, 4. (a) in circle core fiber, and (b) in the ring-core fiber. In the calculation, the signal and pump wavelength is 1550nm and 980nm, respectively. PFM, pump fundamental mode. FRIP, fiber refractive index profile.

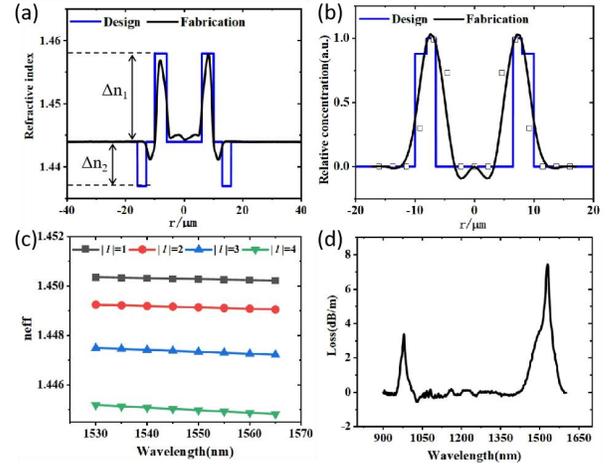

Fig. 2. (a) the design and fabrication of RIP, and (b) EDP(the black hollow square points are the test data, and the black line is the fitting curve), (c) the effective refractive of fabricated fiber indices for the supporting OAM modes as a function of wavelength, (d) the fabricated RC-EDF absorption spectrum.

In this work, utilizing the refractive structure of ring-core with trench assisted and two-layer erbium-doped to promote overlap between excited erbium ions and signal modes, we design and fabricate a RC-EDF for higher-order OAM modes multiplexed amplification with ultra-low DMG, which supports up to the fourth-order OAM mode. A four high order OAM modes ($OAM_{\pm 3,1}$ and $OAM_{\pm 4,1}$) multiplexed amplification with high average gain about 19 dB and an DMG (<1dB) are achieved across the C band.

The RC-EDF incorporating both ring-core with trench outside and the two-layer erbium-doped region, was fabricated by using the combination of modified chemical vapor deposition (MCVD) and plasma chemical vapor deposition (PCVD). The MCVD is used for preparing the refractive index and erbium ions doping in the core region, while the PCVD is used for achieving negative refractive index doping trench which is deployed outside the ring-core. Fig. 2(a) illustrates the designed and fabricated fiber refractive index profile (RIP) of the RC-EDF, where the relative refractive index difference between the step-index ring-core and cladding $\Delta n_1$=0.014, for ensuring that the fiber can support the $OAM_{4,1}$ mode. And the relative refractive index difference between the trench and cladding is set as $\Delta n_2$=-0.007 for fettering the energy distribution of modes into the ring-core further. As shown in Fig. 3(d), the ring-core and trench structure of the fabricated RC-EDF are clearly visible. Under the control of RIP, the energy distribution between the pump mode and signal modes is highly coincident, as shown in Fig. 1(b). Moreover, to reduce DMG effectively, the designed erbium-doped region cover all the core region as shown in Fig. 2(b), where the concentration ratio between the inner and outer layer is 1.22:1 and three boundaries of the two-layer structure lie at the radii of 6.5μm, 8μm and 10μm, respectively. There is high overlap among the modal energy distribution of signal, that of pump and erbium ions, which implies that each signal mode will obtain almost identical stimulated radiation amplification from excited erbium ions. Compare with the large DMG (>5dB) between $OAM_{1,1}$ and $OAM_{2,1}$ in the few-mode EDF with core erbium uniformly doped in Ref. [15], the DMG in our RC-EDF can be expected to be greatly suppressed. This characteristics of RC-EDF will be confirmed by the following experimental results. The fabricated RC-EDF can support OAM modes from 1 to 4 order, and its effective index ($n_{eff}$) curves are calculated based on the measured RIP shown in Fig. 2(c). its small-signal absorption spectrum was measured by the cutback method by using a white light source and an optical spectral analyzer (OSA). The absorption coefficient at 980nm is 3.7dB/m and 1530nm is 8.4dB/m, which is shown in Fig. 2 (d).

To confirm that the third- and fourth-order OAM mode amplification and multiplexed amplification with low DMG can be achieved by this RC-EDF, a core-pumped OAM amplifier testing system was built, which is illustrated in Fig. 3(a). The signal with different order of OAM modes are generated by launching the output of two wavelength tunable lasers through self-made vortex phase plates (VPPs), whose vortex phase are shown in the Fig. 3(b). In Fig. 3(c), clean doughnut shaped intensity profiles and helical interference fringes measured by a charge-coupled device (CCD) before launching into the RC-EDF can be found as expected for generated OAM modes. All the generated OAM modes are combined with three beam splitters (BS). And then the signal modes with the different topology charge are combined with the 976nm single-mode pump by a dichroic mirror (DM). Two additional free-space isolators were incorporated at both ends of the amplifier to prevent the damage of reflected light and undesired parasitic lasing. The coaxial beams (signal and pump) are coupled into the RC-EDF by a 20× objective lens. Note that the output end of the 6.5m RC-EDF was angle-cleaved to further suppress any undesired parasitic lasing while the input end was flat-cleaved for reducing the complexity of the pump and signal coupling [12]. After filtering the residual pump by another DM, he amplified signal would be demodulated coupled into a standard

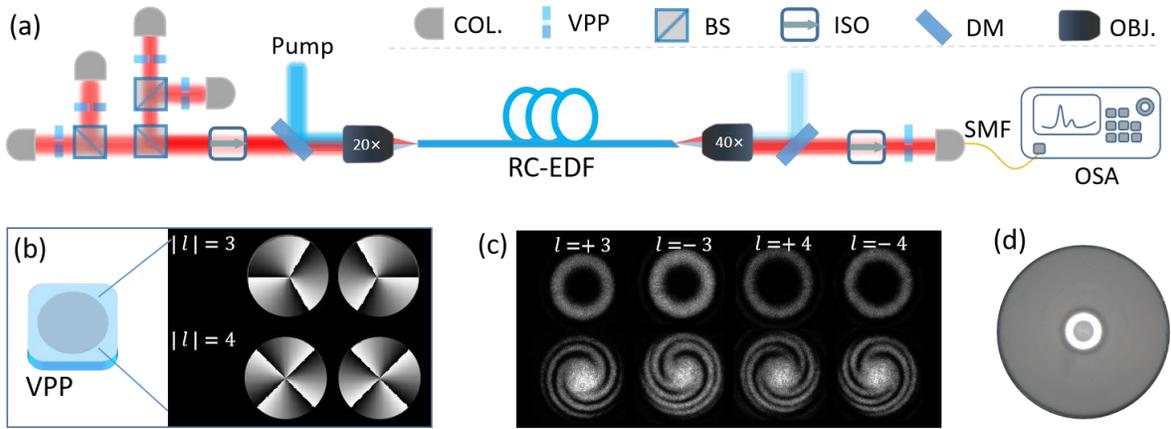

Fig. 3. (a) Experimental setup of the RC-EDF amplifiers. COL, collimator; VPP, vortex phase plate; BS, beam splitter; ISO, isolater; DM, dichroic mirror; OBJ, objective lens; SMF, single-mode fiber; OSA, optical spectral analyzer. (b) the phase of self-made VPP. (c) The intensity distribution and helical interference fringes of the generated OAM modes generated by the self-made VPP. (d) The cross section of the used RC-EDF.

single-mode fiber connected with an OSA for recording the amplified spectrum.

To experimentally testify the OAM modal gain properties, we firstly characterized the modal gain of 4 OAM modes ($l$=+3, -3, +4, -4) by activating separately. In the experiments, the signal power coupling into RC-EDF is about -10dBm after taking the coupled loss of each mode into account. Fig. 4 shows the modal intensity distribution and helical interference fringes of OAM modes amplified individually under 150 mW and 250 mW pump power. The energy distribution and the interference pattern of OAM modes are enhanced with the increase of pump power, indicating that each OAM mode is amplified obviously. After considering various losses from the output of the RC-EDF to the input of SMF attached with OSA, we obtained the corresponding modal gain after varying the pump power, and the results are shown in Fig. 5(a). The gain of each mode increases rapidly with the total pump power at the beginning and start to saturation when the pump is over 500mW (the average modal gain (AMG) is close to 24dB). Additionally, As shown in the solid lines with circles in Fig. 5(a), we can see that two sets of OAM mode with opposite topology charge (±3 or ±4) are highly coincident, which makes the difference gain of them close to zero.

Evaluating the DMG by individually launching different modes has certain rationality [16], however, this method cannot reflect the mode competition of different spatial modes. Therefore, we characterized the gain of 4 different OAM modes multiplexed amplification. In order to distinguish the received power among those 4 OAM modes in the experiments, we first measured the receiving crosstalk (CT) of all modes demultiplexed by a self-made VPP. It should be noticed that the crosstalk includes the crosstalk from the OAM mode multiplexer, de-multiplexer, and the RC-EDF. As shown in Fig. 6 (a), the crosstalk matrix about all multiplexed modes was measured in 500mW total pump power. The CT between 4 multiplexed higher-order OAM modes is less than −9 dB and the lowest CT values are distributed diagonally. To further make sure the signal mode demultiplexed by VPPs, the measured intensity profiles of demodulation of all OAM multiplexed modes in 1550nm wavelength are illustrated in the Fig. 6 (b). One can clearly see that the expected signal mode was transformed back to the Gaussian point while the others were converted to the other order OAM modes. After taking the crosstalk into account in the process of obtaining the mode gain, we got the gain and the DMG curves, as shown in the Fig. 5(b). For those OAM modes, the gain results are also highly consistent. The mode gain and DMG are close to 20dB and 0.55dB, respectively, when the total pump power increases to 500 mW.

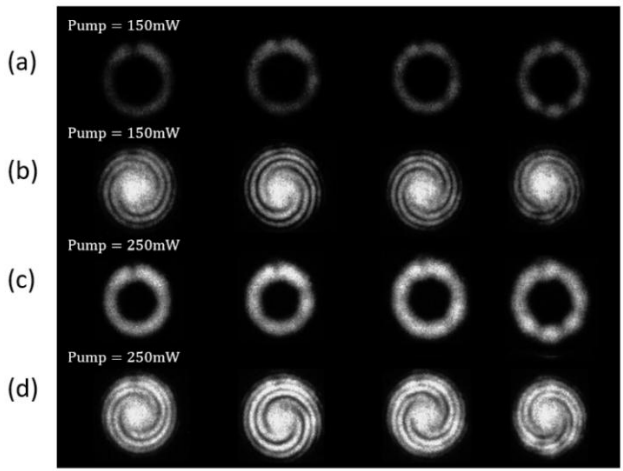

Fig. 4. The modal intensity distribution (a) and (c), and helical interference fringes (b) and (d), of $OAM_{\pm 3,1}$ and $OAM_{\pm 4,1}$ signal mode at the 1550nm wavelength under 150 mW and 250 mW total pump power, respectively.

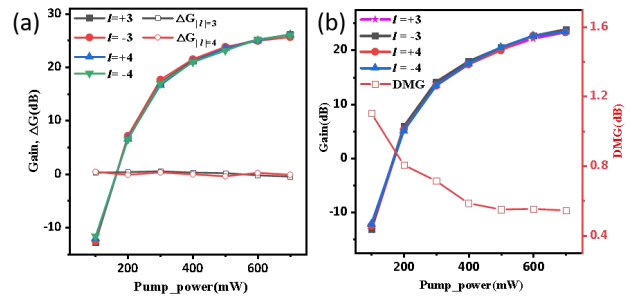

Fig. 5. Measured modal gain in 1550nm for the RC-EDF separately (a) and simultaneously (b), respectively. The solid line is the modal gain of different OAM modes as a function of the pump power; and the dotted line in(a) is the difference gain of two sets of the conjugate OAM modes. the dotted line in (b) is the DMG as a function the pump power.

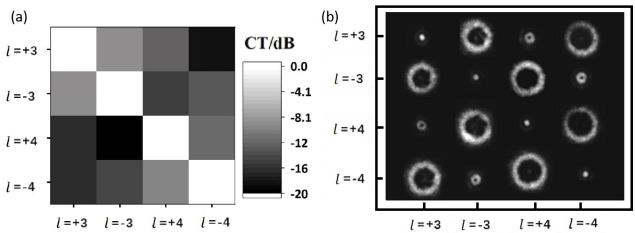

Fig. 6. (a) the image of modal received crosstalk matrix. (b) the intensity of demodulation of all multiplexed OAM modes, respectively.

Finally, the gain spectra of RC-EDF in C-band were measured, and the gain and noise figure (NF) were tested repeatedly by simultaneously launching 4 OAM modes ($l$=+3, -3, +4, -4) with wavelength scanning from 1530m to 1565nm and the total pump power fixing at 500mW. Fig. 8(a) shows that the amplified spectra (measured with an OSA resolution of 0.5 nm) of OAM mode ($l$=-4) as 4 OAM modes multiplexed. It is obvious that the lowest output is located at 1565nm and the highest is laid at 1555nm, which indicates that the highest gain (AMG is 21.1dB) at 1555nm and the lowest gain (AMG is 16.3dB) at 1565nm. Although the gain excursion over the whole C-band for all OAM multiplexed modes is about 5 dB, it is clear that the DMG for each fixed wavelength point is less than 1 dB, and the average modal gain across the C-band is more than 19 dB. As for the NF, the measured NF tends to 3.4 ~5.5 dB across the C band.

In conclusion, we have demonstrated the third- and fourth-order OAM modes amplification separately and simultaneously, by using the RC-EDF we designed and fabricated. This is the first report as far as we know. The measured average gain of 4 modes ($OAM_{\pm3,1}$ and $OAM_{\pm4,1}$) is about 24dB at 1550nm by amplifying separately, and 19 dB by amplifying simultaneously with ultra-low DMG (<1dB) cover C band, and we believe it can achieve higher gain by adjusting the power of pump and signal input[16].

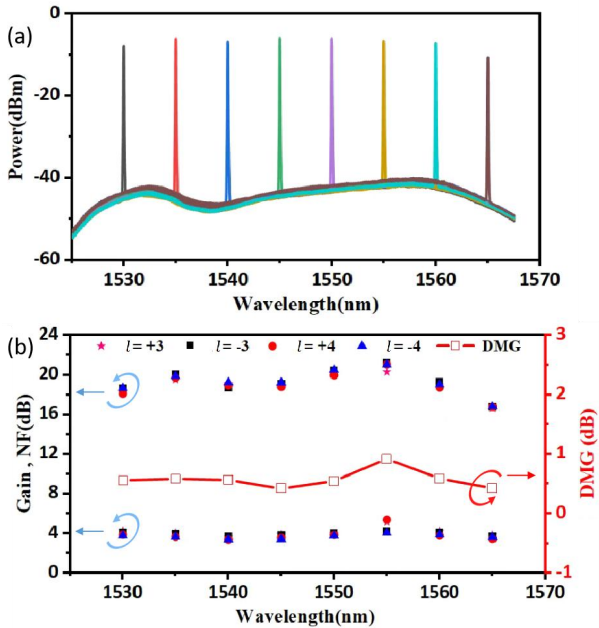

Fig. 9. (a) Measured amplified spectra across C band for OAM mode ($l$=-4) in the case of four OAM modes multiplexed in 500mW pump power. (b) Modal gain, DMG and NF of the RC-EDF amplifier as a function of the C band. In this measurement, the total pump power is fixed at 500mW.

Moreover, the results also show that OAM modes with opposite topology charge obtain the gain is extremely similar whether it is amplified separately or simultaneously. Our experiment lays a foundation for not only the realization of longer distance communication, but also the application in structured light beams trapping and material processing [19]. And we expect that the RC-EDF can be regarded as a good candidate to realize the inline OAM transmission amplification by precisely matching the mode profiles between the RC-EDF and OAM transmission fiber.

**Funding.** This work is funded by the National Key R&D Program of China (2019YFA0706300); National Natural Science Foundation of China (U2001601, 62035018, 61775085, 61875076, 61935013, U1701661); Local Innovation and Research Teams Project of Guangdong Pearl River Talents Program (2017BT01X121); Science and Technology Planning Project of Guangdong Province (2018B010114002 , 2020B0303040001); Natural Science Foundation of Guangdong Province (2021A1515011837)

**Disclosures.** The authors declare no conflicts of interest.